\documentstyle[aps,prl,epsf,floats,12pt]{revtex}

\draft
\begin{document}
\renewcommand{\topfraction}{0.99}

\title{Transformation of Statistics 
       in Fractional Quantum Hall Systems}

\author{
   \underline{John J. Quinn}$^{1}$,
   Arkadiusz W\'ojs$^{1,2}$,
   Jennifer J. Quinn$^{3}$, and
   Arthur T. Benjamin$^{4}$\\[1ex]}
\address{\sl\footnotesize
   $^1$University of Tennessee, Knoxville, Tennessee 37996, USA \\
   $^2$Wroclaw University of Technology, Wroclaw 50-370, Poland \\
   $^3$Occidental College, Los Angeles, California 90041\\
   $^4$Harvey Mudd College, Claremont, California 91711\\[3ex]}
\address{\rm\footnotesize\parbox{6.5in}{
A Fermion to Boson transformation is accomplished by attaching 
to each Fermion a tube carrying a single quantum of flux oriented
opposite to the applied magnetic field.
When the mean field approximation is made in Haldane's spherical
geometry, the Fermion angular momentum $l_F$ is replaced by 
$l_B=l_F-{1\over2}(N-1)$.
The set of allowed total angular momentum multiplets is identical 
in the two different pictures.
The Fermion and Boson energy spectra in the presence of many body 
interactions are identical only if the pseudopotential $V$ 
(interaction energy as a function of pair angular momentum $L_{12}$) 
increases as $L_{12}(L_{12}+1)$.
Similar bands of low energy states occur in the two spectra if $V$ 
increases more quickly than this.\\[1ex]
}
}
\maketitle

\vspace{-1mm}
\subparagraph*{Introduction.}
In two dimensional systems particle statistics can be changed by 
making a Chern--Simons (CS) transformation\cite{wilczek}.
This transformation can be described as attaching to each particle 
an infinitesimal flux tube carrying a fictitious flux $\phi$ and 
a fictitious charge $q$, which couples to the vector potential 
produced by the flux tubes on every other particle in the standard way.
If $q$ is equal to the electron charge and $\phi$ is an even number 
times $\phi_0=hc/e$, the quantum of flux, no change in statistics 
occurs.
If $\phi$ is and odd number times $\phi_0$, Fermions are transformed 
into Bosons.
The ``gauge field'' interactions associated with the fictitious charge 
and vector potential produced by the fictitious flux make the Hamiltonian 
of the system considerably more complicated.
Only when the mean field approximation is made does the problem simplify.
Jain\cite{jain} introduced the mean field CS transformation to give 
a simple intuitive picture of the hierarchy of fractional quantum Hall 
(FQH) states in terms of the resulting composite Fermions (CF).
Shortly after the introduction of the CF picture, Xie et al.\cite{xie}
introduced the Fermion--Boson mapping connecting a 2D Fermion system 
at filling factor $\nu_F$ to a 2D Boson system with filling factor 
$\nu_B=\nu_F(1-\nu_F)^{-1}$.
These authors noted that the size of the Hilbert space for the Fermion
and Boson systems was identical, and they found that the F$\rightarrow$B
mapping accurately transformed the ground state of the Fermion system 
into that of the Boson system only if these ground states were 
incompressible FQH states.
In this paper we show that the F$\rightarrow$B transformation leads 
to identical energy spectra if and only if the pseudopotential 
$V(L_{12})$ describing the interaction energy as a function of the 
pair angular momentum $L_{12}$ is of the ``harmonic'' form 
$V_H(L_{12})=A+BL_{12}(L_{12}+1)$ where $A$ and $B$ are constants.
Laughlin\cite{laughlin} correlations occur when the actual 
pseudopotential $V(L_{12})$ rises more quickly with increasing 
$L_{12}$ than $V_H(L_{12})$.
Anharmonic effects (due to $\Delta V(L_{12})=V(L_{12})-V_H(L_{12})$) 
cause the interacting Fermion and interacting Boson spectra to differ 
for every value of the filling factor.

It is well known that in the Haldane spherical geometry\cite{haldane} 
the mean field CF transformation changes an electron of angular 
momentum $l_F$ to a CF of angular momentum $l_F^*=|l_F-p(N-1)|$
\cite{chen}.
Here $l_F=S_F$, one half the strength (measured in units of
$\phi_0$) of the magnetic monopole which produces the radial magnetic 
field $B=2S_F\phi_0(4\pi R^2)^{-1}$ at the spherical surface of radius
$R$ on which the $N$ electrons are confined, and $p$ is an integer.
For an $N$ Boson system, the composite Boson transformation replaces 
$l_B$ by $l_B^*=|l_B-p(N-1)|$.
In the F$\rightarrow$B mapping $l_F$ is replaced by 
$l_B=|l_F-{1\over2}(N-1)|$.

\vspace{-4mm}
\subparagraph*{Angular Momentum Addition: Useful Theorems.}
When a shell of angular momentum $l$ contains $N$ identical particles
(Fermions or Bosons), the resulting $N$ particle states can be 
classified by eigenvectors $\left|L,M,\alpha\right>$, where $L$ 
is the total angular momentum, $M$ its $z$-component, and $\alpha$ 
a label which distinguishes independent multiplets with the same $L$.
In the mean field CF (CB) transformation $l_F$ ($l_B$) is transformed
to $l_F^*$ ($l_B^*$).
In trying to understand why the mean field CF picture correctly 
predicted the low lying band of states in the electron spectrum, 
several simple conjectures were proposed on the basis of numerical 
studies of a finite number of particles\cite{arek1}.
These conjectures have been elevated to the status of theorems by 
rigorous mathematical proof\cite{quinn} using partition theory.

\paragraph*{Theorem 1.}
The set of allowed total angular momentum multiplets of $N$ Fermions
each with angular momentum $l_F^*$ is a subset of the set of allowed
multiplets of $N$ Fermions each with angular momentum $l_F=l_F^*+(N-1)$.

\paragraph*{Theorem 2.}
The set of allowed total angular momentum multiplets of $N$ Bosons 
each with angular momentum $l_B$ is identical to the set of multiplets 
for $N$ Fermions each with angular momentum $l_F=l_B+{1\over2}(N-1)$.

From Theorem 2 it follows immediately that Theorem 1 also applies 
to Bosons.
Theorem 2 is a stronger statement than a simple equality of sizes
of the many body Hilbert spaces\cite{xie}.

\vspace{-4mm}
\subparagraph*{Interaction Effects.}
It has been shown that for the harmonic pseudopotential $V_H(L_{12})$ 
the energy of any multiplet of angular momentum $L$ is given by
$E_{\alpha}(L)=A\times {1\over2}N(N-1)+B\times\{N(N-2)l(l+1)+L(L+1)\}$.
The energy is independent of $\alpha$, so that every multiplet with the 
same value of $L$ has the same energy.
If $B_F=B_B=B$, then the spectrum of $N$ Bosons each with angular 
momentum $l_B$ is identical (up to a constant) to that of $N$ Fermions 
each with angular momentum $l_F=l_B+{1\over2}(N-1)$.
This is a rather surprising result because Fermions and Bosons sample
different sets of values of the pair angular momentum.
For example, for $N=9$ and $l_F=12$ (corresponding to $\nu_F={1\over3}$) 
the allowed values of the Fermion pair angular momentum consist of all 
odd integers between 1 and 23; for the corresponding Boson system with 
$l_B=8$ ($\nu_B={1\over2}$), the allowed values of $L_{12}$ are all 
even integers between 0 and 16.

Xie et al.\cite{xie} determined the Boson and Fermion eigenfunctions
by exact numerical diagonalization for six particle systems connected 
through the F$\rightarrow$B transformation.
They then transformed the Boson eigenfunctions into Fermion 
eigenfunctions by multiplying them by $\prod_{i<j}(z_i-z_j)$, 
as required by the B$\rightarrow$F transformation.
The overlap of these transformed Boson eigenfunctions with the exact
Fermion eigenfunctions was then evaluated.
The overlap was close to unity for incompressible quantum fluid
states when the full Coulomb interaction was used, but it was considerably
smaller when certain model pseudopotentials were used.

\begin{figure}[t]
\epsfxsize=6.5in
\epsffile{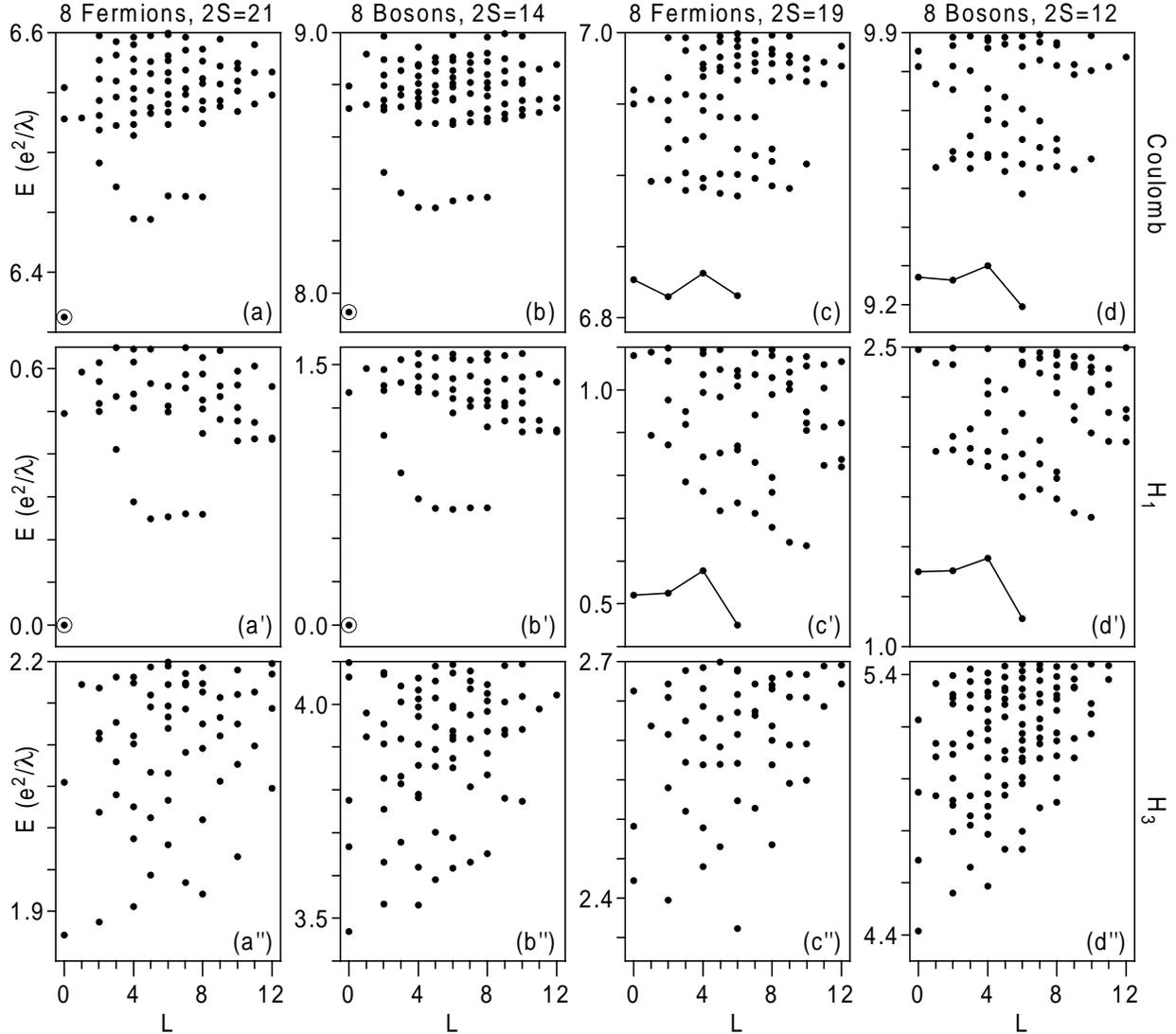}
\caption{
   The energy spectra (energy $E$ vs.\ angular momentum $L$) 
   of eight Fermions at the monopole strength $2S_F=21$ 
   (filling factor $\nu_F={1\over3}$; circles) and 19 
   (two Laughlin quasielectrons; lines), and of eight Bosons 
   at $2S_B=14$ ($\nu_B={1\over2}$; circles) and 12 (two 
   quasielectrons; lines) for the Coulomb pseudopotential 
   in the lowest Landau level (a--d), and for the model 
   pseudopotentials $H_1$ (a$'$--d$'$), and $H_3$ 
   (a$''$--d$''$). 
   $\lambda$ is the magnetic length.
}
\label{fig1}
\end{figure}

We have evaluated numerically the eigenstates of an eight electron 
system at $2S_F=19$ to 23 (these states correspond to Laughlin 
$\nu_F={1\over3}$ states with zero, one, or two QP's) for a number 
of different pseudopotentials.
In frames (a--a$''$) and (b--b$''$) of Fig.~\ref{fig1} we contrast the 
energy spectra for the Fermion and Boson systems at $\nu_F={1\over3}$ 
($\nu_B={1\over2}$) for the Coulomb pseudopotential appropriate for the 
lowest Landau level and for the model pseudopotentials $H_1$ and $H_3$.
$H_1$ is defined to have only the largest pseudopotential coefficient
[$V(L_{12}=2l_F-1)$ for Fermions and $V(L_{12}=2l_B)$ for Bosons] equal
to its Coulomb value and all other pseudopotential coefficients equal
to zero.
$H_3$ has the two largest pseudopotential coefficients equal to their
Coulomb values and all other coefficients equal to zero.
In frames (c--c$''$) and (d--d$''$) we do the same for the state 
containing two Laughlin quasielectrons (QE).
The lowest states in (a, a$'$, b, b$'$) are quite similar consisting 
of a Laughlin $L=0$ ground state.
The magnetoroton band (at $2\le L\le8$) is apparent in all four 
spectra, although the gaps and band widths are different for different 
pseudopotentials.
The lowest states in (c, c$'$, d, d$'$) are also similar containing
two QE's with $l_{\rm QE}={1\over2}(N-1)={7\over2}$ giving $L=N-2$, 
$N-4$, $\dots=0$, 2, 4, and 6.
The pseudopotential $H_3$ used in (a$''$--d$''$) gives very different 
results.
We believe that this behavior results because the pseudopotential 
used in $H_3$ is not ``short range'' in the sense that it does not 
increase faster than $V_H(L_{12}))$ in the region of non-vanishing 
$V(L_{12})$.
This behavior of the pseudopotential results in correlations that 
are very different from Laughlin correlations, and it accounts for 
the poor overlap found by Xie et al.\cite{xie} for certain model 
pseudopotentials.

\vspace{-4mm}
\subparagraph*{Quasiparticles.}
The F$\rightarrow$B transformation allows us to better understand 
the Boson\cite{haldane} vs.\ Fermion\cite{sitko,arek2} description 
of QP's in incompressible FQH states.
Laughlin condensed states having $\nu_F=(2p+1)^{-1}$ (where $p$ is a 
positive integer) occur at $2S_F=(2p+1)(N-1)$ in the Haldane spherical
geometry.
The CF transformation gives an effective angular momentum
$l_F^*=S_F^*=S-p(N-1)={1\over2}(N-1)$ when $2p$ flux quanta are attached 
to each electron.
Thus the $N$ CF's fill the $2l^*+1$ states of the lowest CF shell giving 
an $L=0$ incompressible ground state.

The F$\rightarrow$B transformation gives $2S_B=2S_F-(N-1)=2p(N-1)$
and a Boson filling factor of $\nu_B=(2p)^{-1}$.
Making a CB transformation gives $l_B^*=S_B^*=S_B-p(N-1)=0$.
This also gives an $L=0$ incompressible ground state because each
CB has $l_B^*=0$.
Thus the CF description of a Laughlin state has one filled CF shell
of angular momentum $l_F^*={1\over2}(N-1)$, while the CB description
has $N$ CB's each with angular momentum $l_B^*=0$.

For $2S_B=2p(N-1)\pm n_{\rm QP}$, where the $+$ and $-$ occur for
quasiholes (QH) and quasielectrons (QE), respectively, we define 
$2l_B^*=|2S_B^*|=n_{\rm QP}$.
This gives exactly the same set of angular momentum multiplets as
obtained in the CF picture with $2S_F=(2p+1)(N-1)\pm n_{\rm QP}$
only if a hard core repulsion
forbids the Boson QE pair from having the largest allowed pair 
angular momentum $L_{12}^{\rm MAX}=N$\cite{he}.
This behavior is observed in Fig.~\ref{fig1}(c--c$'$, d--d$'$) 
where the Boson treatment of two QE's 
predicts states at $L=0$, 2, 4, 6, and 8, but the $L=8$ state 
does not occur in the low energy band.
Since the description of CB's (with hard core QE interaction) and 
CF's give identical sets of QP states, filled QP levels (implying 
daughter states) occur at identical values of the applied magnetic 
field.

\vspace{-4mm}
\subparagraph*{Acknowledgment.}
John J. Quinn acknowledges partial support from the Materials 
Research Program of Basic Energy Sciences, US Department of Energy.

\end{document}